\documentclass[aps, a4paper,superscriptaddress, nofootinbib, showpacs, twocolumn, 10pt]{revtex4}
\usepackage{amsmath,amssymb,bm,natbib}
\usepackage{epsfig}
\usepackage{graphicx}
\usepackage{slashed}

\newcommand{\beq}{\begin{equation}}
\newcommand{\eeq}{\end{equation}}
\newcommand{\bea}{\begin{eqnarray}}
\newcommand{\eea}{\end{eqnarray}}
\newcommand{\ba}{\begin{align}}
\newcommand{\ea}{\end{align}}
\newcommand{\bfig}{\begin{figure}}
\newcommand{\efig}{\end{figure}}

\newcommand{\D}{\displaystyle}

\newcommand{\gev}{\, \text{GeV}}
\newcommand{\mev}{\, \text{MeV}}

\newcommand{\tin}{t_{\rm in}}
\newcommand{\la}{\langle}
\newcommand{\ra}{\rangle}

\newcommand{\omnes}{{\cal{O}}}

\begin{document}

\vspace{1cm}

\title{Implications of the recent high statistics determination of
  the pion electromagnetic form factor in the timelike region}
\author{B.Ananthanarayan}
\affiliation{Centre for High Energy Physics,
Indian Institute of Science, Bangalore 560 012, India}
\author{Irinel Caprini}
\affiliation{National Institute of Physics and Nuclear Engineering\\
POB MG 6, Bucharest, R-76900, Romania}
\author{I. Sentitemsu Imsong}
\affiliation{Centre for High Energy Physics,
Indian Institute of Science, Bangalore 560 012, India}

~\vspace{0.5cm}

\begin{abstract}
 The recently evaluated two-pion contribution to the muon $g-2$ and the phase of the pion electromagnetic form factor in 
the elastic region, known from  $\pi\pi$ scattering by Fermi-Watson theorem, are exploited by analytic techniques for 
finding correlations between  the coefficients of the Taylor expansion at $t=0$ and the values of the form factor at 
several points in the spacelike region. We do not use specific parametrizations and the results are fully independent 
of the unknown phase in the inelastic region. Using for instance, from recent determinations, 
$\la r^2_\pi \ra = (0.435 \pm 0.005) \,{\rm fm}^2$ and  $F(-1.6\, {\rm GeV}^2)=0.243^{+0.022}_{-0.014} $,
we obtain the allowed ranges $3.75 \gev^{-4}\lesssim c \lesssim 3.98\gev^{-4}$ and 
$9.91 \gev^{-6}\lesssim d \lesssim 10.46 \gev^{-6}$ for the curvature and the next Taylor coefficient, with a strong
correlation between them. We also predict a large region in the complex plane where the form factor cannot have zeros.
\end{abstract}

\maketitle
\section{Introduction}
\label{sec:intro}
The pion electromagnetic form factor $F(t)$, defined by the matrix element
\begin{eqnarray}\label{defe} 
 \langle \pi^+(p')\vert J_\mu^{\rm elm} \vert
\pi^+(p)\rangle= (p+p')_\mu F(t)
\end{eqnarray}
where  $q=p-p'$ and $t=q^2$,  plays a central role in strong interaction 
dynamics.  From the general principles of quantum field theory, it follows that $F(t)$ is  normalized to
 $F(0)=1$ and is a real analytic function in the $t$-plane cut along the real axis from the unitarity 
threshold $t_+=4 M_\pi^2$ to infinity. At low energies its properties are described by chiral perturbation 
theory (ChPT), the low energy effective theory of the strong interactions 
\cite{Wein,GaLe1}, calculations of the pion form factor being available in ChPT up to two loops 
\cite{GaMe}-\cite{BiTa}. Lattice gauge theory has recently become another useful tool for the 
calculation of the form factor at low energies \cite{Aoki}.  
On the other hand, perturbative  QCD  predicts the behavior at large momenta along the spacelike axis, 
where  $Q^2\equiv -t>0$. The leading order (LO) asymptotic term is \cite{Farrar:1979aw}-\cite{pQCD4}
\beq\label{eq:qcd}
F (-Q^2)\sim \frac{16 \pi F_\pi^2 \alpha_s(Q^2)}{Q^2}, \quad\quad Q^2\to \infty,
\eeq
where $F_\pi$ is the pion decay constant and $\alpha_s(Q^2)=4\pi/[9\ln(Q^2/\Lambda^2)]$ is the  running strong 
coupling to one loop. Next-to-leading-order corrections to (\ref{eq:qcd}) were calculated by various groups  \cite{Melic}-\cite{Chen}. 
As discussed, for instance in \cite{pQCD2,DoNa,Leutwyler:2002}, the transition to the perturbative QCD regime  seems to occur quite slowly in this case.

The experimental information available on the   pion form factor is very rich.   This quantity was measured at spacelike
values $Q^2>0$  with increasing precision from electron-pion scattering and pion electroproduction from nucleons 
\cite{Bebek}-\cite{Huber}. On the timelike cut, where the form factor is complex,  the Fermi-Watson theorem implies that 
in the elastic region its phase is equal to the phase-shift of the $P$-wave of the $\pi\pi$ amplitude, calculated 
recently with precision using Roy equations and fixed-t dispersion relations \cite{ACGL,CGL,KPY}.
The modulus has been measured from the cross section of $e^+e^-\to\pi^+\pi^-$ by several groups in the 
past \cite{OLYA1}-\cite{DM2}, and more recently to high accuracy by the BABAR \cite{BABAR} and
 KLOE \cite{KLOE1,KLOE2} collaborations.  These data have been used for an accurate evaluation 
of the two-pion contribution to the muon anomalous magnetic moment \cite{Davier:2009, Davier:2010}.

 The constraints imposed on the pion form factor by analyticity and unitarity  have been exploited in many 
works (the list \cite{Raszillier}-\cite{BDDL} covers only partly a very rich literature). Different 
analytic representations, either as standard dispersion relations \cite{DoNa},  phase (Omn\`es-type)  
\cite{LaSt,GuPi,PiPo,Geshkenbein,Leutwyler:2002,Guo} or modulus \cite{Geshkenbein} representations, 
as well as expansions based on conformal mappings \cite{LaSt,BuLe,Leutwyler:2002}  or Pad\'e-type 
approximants \cite{Masjuan}, have been constructed in order to correlate the low- and high-energy 
properties of the form factor. Of special interest is the issue of the zeros of the form factor, 
investigated by means of dispersive sum-rules \cite{Cronstrom,DuMe,Shcherbin,Geshkenbein,Leutwyler:2002} 
or by the more powerful  techniques of analytic optimization theory \cite{Raszillier,RaSc,RaSc2}.   
In \cite{IC, AnRa1, AnRa2, Abbas:2009} similar functional-analytic techniques were applied for 
deriving bounds on the expansion coefficients at $t=0$, from  an weighted integral of the modulus 
squared along the cut, known from unitarity and dispersion relations for a related QCD correlator.

In the present paper we address the same problem, {\em i.e.} to find constraints on the coefficients 
appearing in the Taylor expansion 
\begin{equation}\label{eq:taylor}
	F(t) = 1 + \D\frac{1}{6} \la r^2_\pi \ra t + c t^2 + d t^3 + \cdots
\end{equation}
from a well-defined input on the timelike axis, and also include information coming from high precision
experiments that measure the form factor in the spacelike region. We also consider the problem of the zeros, 
and obtain a region in the complex $t$-plane where zeros are excluded. The main reason of 
revisiting the problem is the recent high statistics measurement of the modulus $|F(t)|$ on 
the unitarity cut by BABAR \cite{BABAR} and KLOE  \cite{KLOE1,KLOE2} experiments.  As we will show, 
this  information leads to stringent constraints,  of a remarkable level for a prediction independent 
of any specific parametrization. 

We apply a technique discussed in \cite{IC,Abbas:2010EPJA}, which makes use of information on both 
the phase and modulus, and was shown recently \cite{Abbas:2010EPJAL, Abbas:2010PRD} to place stringent 
bounds on the  $K\pi$ weak form factors.  As first input we use the Fermi-Watson theorem, 
according to which one has, modulo $\pi$,
\beq\label{eq:watson}
{\rm Arg} [F(t+i\epsilon)]=\delta_1^1(t), \quad\quad t_+<t<\tin,
\eeq
where $\delta_1^1(t)$ is the phase-shift of the $P$-wave of $\pi\pi$ elastic scattering and 
$\tin$ the first inelastic threshold. As discussed previously \cite{LaSt,Leutwyler:2002}, 
inelasticity in the case of the pion vector form factor is negligible below the opening of 
$\pi\omega$ channel, so we take $\tin=(M_\pi+M_\omega)^2$. Below this energy, the phase 
$\delta_1^1(t)$ is known with precision from Roy equations and fixed-t dispersion relations 
for $\pi\pi$ scattering \cite{ACGL,CGL,KPY}.

We also include information on the modulus, generically expressed by an integral relation
\beq\label{eq:L2}
\D\frac{1}{\pi} \int_{\tin}^{\infty} dt 
\rho(t) |F(t)|^2 \leq  I,
\eeq
where $\rho(t)$ is a positive definite weight in the region of integration and $I$ is a known quantity.
Actually,  (\ref{eq:L2}) does not fully account for the present information on $|F(t)|$:  
indeed, except for a small region near the threshold $t_+=4 M_\pi^2$, the modulus is measured also 
below the inelastic threshold $\tin$, i.e $|F(t)|$ is measured more or less pointwise, 
at every $t$, not only in averaged form as in (\ref{eq:L2}). In principle, the accurate 
knowledge of the phase and modulus on a region on the unitarity cut is sufficient  to 
pin down the  form factor everywhere due to analyticity.  In practice, however,   due to 
the well-known``instability" of analytic continuation, the uncertainties, however small, lead 
to  solutions which are very different at points outside the original data interval. Therefore, 
we do not proceed by constructing parametrizations of the form factor on the timelike axis, but 
consider instead the global class of functions compatible with the adopted input, and derive 
constraints on various quantities of interest from this class of functions.  As we shall see, 
even the input  (\ref{eq:L2}) leads to quite strong constraints on the properties of the 
form factor near $t=0$ and in the complex plane. Thus the chosen method is fully justified by the results
that have been obtained.

A further open point is the choice of the weight $\rho(t)$ in  (\ref{eq:L2}). In principle, a large 
class of positive weights, leading to a convergent integral for $|F(t)|$ compatible with the asymptotic 
behavior (\ref{eq:qcd}), can be adopted. The optimal procedure is to vary $\rho(t)$ over 
a suitable admissible class and take the best result. This approach will be investigated in a future work. 
In the present paper we make the particular choice  that corresponds to the two-pion contribution to the muon $g-2$, when 
the weight $\rho(t)$ has the form
\bea\label{eq:amu}
\rho(t) = \frac{\alpha^2 M_{\mu}^2}{12 \pi} \frac{(t - t_+)^{3/2}}{t^{7/2}} K(t), \nonumber\\
K(t) = \int_0^1 du\, \frac{(1-u) u^2} {1- u + M_\mu^2 u^2/t},
\eea
and the right-hand side (rhs) of (\ref{eq:L2}) is the two-pion contribution to the muon anomaly in the range $t>\tin$,
\beq\label{eq:amuI}
I= \hat a_\mu^{\pi\pi}.
\eeq 
The practical motivation of this  particular choice is that an accurate evaluation of the  two-pion 
contribution to the muon anomaly, taking into account the correlations between different points, 
is available from the refs. \cite{Davier:2009,Davier:2010}. As a result, this choice guarantees a very precise input.  
We must emphasize that, once the input (\ref{eq:watson})-(\ref{eq:amuI}) is adopted, the treatment is 
optimal and no information is lost. A posteriori, it turns out that the results given by this choice are 
quite stringent.

In addition to the above input from the timelike axis, we include the values of $F(t)$ measured 
experimentally at some spacelike points
\beq\label{eq:spacelike}
F(t_n)= f_n \pm \delta f_n, \quad\quad t_n<0, \quad n=1,...., N,
\eeq
where we use the most recent high precision experimental information from \cite{Horn,Huber}.
Thus, we will be employing as input Eqs. (\ref{eq:watson})-(\ref{eq:spacelike}) in order to obtain 
correlations between the coefficients of the Taylor expansion (\ref{eq:taylor}). We will investigate 
also the issue of the possible zeros of the form factor, deriving regions where zeros are forbidden.
 
In Sec. \ref{sec:method}  we briefly review the mathematical method and in Sec. \ref{sec:exp} the 
experimental information that goes into 
our computation. In Sec. \ref{sec:cd}, we present our results for the parameters 
$(c,d)$ and compare them with results available in the literature. In Sec. \ref{sec:zeros} we derive 
regions  where zeros are excluded along
the real axis and in the complex $t$-plane, and in Sec. \ref{sec:conclusion} some discussions
and our conclusions are presented.

\section{Basic formulae}\label{sec:method}
For solving the problem we follow a mathematical method presented in \cite{IC,Abbas:2010EPJA}.
 We first define the Omn\`{e}s function
\beq	\label{eq:omnes}
 \omnes(t) = \exp \left(\D\frac {t} {\pi} \int^{\infty}_{t_+} dt' 
\D\frac{\delta (t^\prime)} {t^\prime (t^\prime -t)}\right),
\eeq
where $\delta(t)=\delta_1^1(t)$   for 
$t\le \tin$, and is an arbitrary function, sufficiently  smooth ({\em i.e.}
Lipschitz continuous) for $t>\tin$. As shown in  \cite{Abbas:2010EPJA}, the results do not depend on the 
choice of the function  $\delta(t)$ for $t>\tin$.
A crucial remark is that the function $h(t)$ defined by
\beq\label{eq:h}
F(t)=\omnes(t) h(t)
\eeq
is analytic in the $t$-plane cut only for $t>\tin$. 
The equality (\ref{eq:L2}), written in terms of $h(t)$ as
\beq\label{eq:hL2}
\D\frac{1}{\pi} \int_{\tin}^{\infty} dt\, 
\rho(t) |\omnes(t)|^2 |h(t)|^2 =  \hat{a}^{\pi\pi}_\mu,
\eeq
can be expressed in a canonical form, if we perform the conformal transformation
\beq\label{eq:ztin}
\tilde z(t) = \frac{\sqrt{\tin} - \sqrt {\tin -t}} {\sqrt{\tin} + \sqrt {\tin -t}}\,,
\eeq
which maps the complex $t$-plane cut for $t>\tin$  onto the unit disk $|z|<1$ in the $z$-plane defined 
by $z\equiv\tilde z(t)$,  and define a function $g(z)$ analytic in   $|z|<1$ by
\beq\label{eq:gF}
 g(z) = w(z)\, \omega(z) \,F(\tilde t(z)) \,[\omnes(\tilde t(z)) ]^{-1}.
\eeq 
In this relation  $\tilde t(z)$ is the inverse of $z = \tilde z(t)$, for $\tilde z(t)$ as defined in
(\ref{eq:ztin}), and the last two factors give the function $h(\tilde t(z))$ defined in (\ref{eq:h}), 
which is analytic in 
$|z|<1$.
Finally, $w(z)$ and $\omega(z)$ are outer functions, {\it i.e.} functions analytic and without zeros in
$|z|<1$, defined in terms of their modulus on the boundary, related to  $\sqrt{\rho(t)}$ and  $|\omnes(t)|$, respectively. Equivalent integral representations of the outer functions in terms of their modulus can be written either in the $z$ or $t$ variables.  In particular, we use 
\beq\label{eq:wgen}
w(z)=\exp\left[\frac{1}{2\pi} \int_{0}^{2\pi} {\rm d}\theta \,
\frac{\zeta+z}{\zeta-z}\,\ln |w(\zeta)| \right],\quad \zeta=\exp(i\theta),
\eeq 
where 
\beq\label{eq:wrho}
|w(\zeta)|^2=  \rho(\tilde t(\zeta))\, \left|\frac{{\rm d} \tilde t(\zeta)}{{\rm d}\zeta}\right|,
\eeq
and 
\beq\label{eq:omega}
 \omega(z) =  \exp \left(\D\frac {\sqrt {\tin - \tilde t(z)}} {\pi} \int^{\infty}_{\tin}  \D\frac {\ln |\omnes(t^\prime)|\, {\rm d}t^\prime}
 {\sqrt {t^\prime - \tin} (t^\prime -\tilde t(z))} \right).
\eeq 
 Then (\ref{eq:hL2}) can be written as
\beq\label{eq:gI1}
\frac{1}{2 \pi} \int^{2\pi}_{0} {\rm d} \theta |g(\zeta)|^2 = \hat a_\mu^{\pi\pi}.
\eeq
>From (\ref{eq:ztin}) it follows that the origin $t=0$ of the $t$-plane is mapped onto the origin $z=0$ of the $z$-plane. 
Therefore, from (\ref{eq:gF}) it follows that each coefficient $g_k \in R$ of the expansion
\begin{equation}\label{eq:gz}
g(z)=g_0+ g_1 z+ g_2 z^2 +  g_3 z^3+\dots
\end{equation}
is expressed in terms of the coefficients of order lower or equal to $k$,  of the Taylor expansion (\ref{eq:taylor}). 
Moreover, the values $F(t_n)$ of the form factor at a set of real points
$t_n<0,\, n=1,2,..., N$, lead to the values
\beq\label{xin}
g(z_n)=w(z_n)\, \omega(z_n) \,F(t_n) \,[\omnes(t_n) ]^{-1}, \quad z_n=\tilde z(t_n).
\eeq
 Then the $L^2$ norm condition (\ref{eq:gI1}) implies the determinantal inequality (for a proof and older 
references see \cite{Abbas:2010EPJA}):
\beq\label{eq:det}
\left|
\begin{array}{c c c c c c}
\bar{I} & \bar{\xi}_{1} & \bar{\xi}_{2} & \cdots & \bar{\xi}_{N}\\	
	\bar{\xi}_{1} & \D \frac{z^{2K}_{1}}{1-z^{2}_1} & \D
\frac{(z_1z_2)^K}{1-z_1z_2} & \cdots & \D \frac{(z_1z_N)^K}{1-z_1z_N} \\
	\bar{\xi}_{2} & \D \frac{(z_1 z_2)^{K}}{1-z_1 z_2} & 
\D \frac{(z_2)^{2K}}{1-z_2^2} &  \cdots & \D \frac{(z_2z_N)^K}{1-z_2z_N} \\
	\vdots & \vdots & \vdots & \vdots &  \vdots \\
	\bar{\xi}_N & \D \frac{(z_1 z_N)^K}{1-z_1 z_N} & 
\D \frac{(z_2 z_N)^K}{1-z_2 z_N} & \cdots & \D \frac{z_N^{2K}}{1-z_N^2} \\
	\end{array}\right| \ge 0,
\eeq
where $K\ge 1$ is an arbitrary integer and 
\beq\label{eq:barxi}
 \bar{I} = \hat{a}^{\pi\pi}_\mu - \sum_{k = 0}^{K-1} g_k^2, \quad  \quad \bar{\xi}_n = g(z_n) - \sum_{k=0}^{K-1}g_k z_n^k.
\eeq
The same relation (\ref{eq:det}) holds  if we replace $\hat a_\mu^{\pi\pi}$ by an upper bound of this 
quantity and the equality sign in (\ref{eq:gI1}) by the $\leq$ sign. Moreover, as shown in \cite{Abbas:2010EPJA}, 
the results depend in a monotonic way on the value of the rhs of (\ref{eq:gI1}), becoming weaker when this 
value is increased. 

The extension to  the case of complex
points $t_n$, which enters in pairs since $F(t^*)=F^*(t)$, is straightforward and will be discussed in Sec. \ref{sec:zeros}.

\section{Experimental input}\label{sec:exp}
We take $\sqrt{\tin}= 0.917\, \gev $,  which corresponds to the first important inelastic threshold, 
due to the $\omega\pi$ pair. The choice of a lower value of $\tin$ is legitimate in the present formalism, 
and we will work  also with $\sqrt{\tin} =0.8\, \gev $, which will allow us to compare the constraining 
power of the input conditions (\ref{eq:watson}) and (\ref{eq:L2}).  

Very precise parametrizations of the phase-shift  $\delta_1^1$ are given in \cite{ACGL, KPY}. 
We use as phenomenological input the phase parametrized as \cite{KPY}
\beq\label{eq:delta11}
{\rm cot}\,\delta_1^1(t)=\frac{\sqrt{t}}{2k_\pi^3} (M_\rho^2 - t) \left(\frac{2 M_\pi^3}{M_\rho^2 \sqrt{t}} + B_0 + 
      B_1 \frac{\sqrt{t} -\sqrt{t_{0} - t}}{\sqrt{t} + \sqrt{t_{0} - t}}\right),\quad 
\eeq
where $k_\pi=\sqrt{t/4 -M_\pi^2}$ and 
\bea
 \sqrt{t_0}=1.05\, \gev,\quad& M_\rho= 773.6 \pm 0.9\, \mev,\, \nonumber\\
 B_0=1.055\pm 0.011,\quad& B_1=0.15 \pm 0.05.
\eea
 The function  $\delta_1^1$ obtained from (\ref{eq:delta11}) is practically identical with the
 phase-shift  obtained in \cite{ACGL}  from Roy equations for $\sqrt{t}\leq 0.8\,\gev$. The uncertainty 
is very small and we have checked that the results are practically insensitive to the variation of the phase-shift 
within the errors.

Above $\tin$ we use in (\ref{eq:omnes}) a smooth phase $\delta(t)$, which approaches asymptotically $\pi$. 
As shown in \cite{Abbas:2010EPJA}, the dependence on $\delta(t)$ of the functions $\omnes$ and $\omega$,  
defined in (\ref{eq:omnes}) and (\ref{eq:omega}) respectively, exactly compensate  each other, leading to 
results fully independent of the unknown phase in the inelastic region.

The two-pion contribution to muon anomaly was evaluated to great precision in  \cite{Davier:2009,Davier:2010}.
The most recent evaluation \cite{Davier:2010}, based on all the available experimental data,
 gives for the total $\pi^{+}\pi^{-}$  contribution to muon anomaly 
the value  $a^{\pi\pi}_\mu = (507.80 \pm 1.22 \pm  2.50 \pm 0.56) \times 10^{-10}$. In our method we need the 
specific contribution  $\hat a^{\pi\pi}_\mu$ of the energies from $\sqrt{\tin}$ to infinity. The values given below\footnote{We are grateful to Bogdan Malaescu for providing us these numbers.} are based on the BABAR data \cite{BABAR}, whose spectrum extends up to 3 GeV. 

For the interval 0.917 - 3 GeV the  two-pion contribution is   $(21.73 \pm 0.24) \times 10^{-10}$.  Increasing the central value by the error, and adding an estimate of about $0.2 \times 10^{-10}$ for the interval from 3 GeV to
$\infty$, gives the close upper bound $\hat a^{\pi\pi}_\mu\leq 22.17 \times 10^{-10}$ 
for the two-pion contribution from 0.917 GeV to $\infty$. As mentioned above,  if we 
use in (\ref{eq:gI1}), instead of the exact value of $\hat a^{\pi\pi}_\mu$ an upper bound 
on this quantity, the results are still valid, but are weaker.  In order to obtain 
results which are in the same time unbiased and stringent,  we need a conservative and 
accurate estimate of $\hat a^{\pi\pi}_\mu$.

For the interval 0.8 - 3 GeV the  two-pion contribution in \cite{Davier:2010} is 
$(94.25 \pm  0.77) \times 10^{-10}$. Increasing as before the central value by the error, 
and adding $0.2 \times 10^{-10}$ for the interval from 3 GeV to $\infty$, we obtain   
$\hat a^{\pi\pi}_\mu\leq  95.23 \times 10^{-10}$ for the two-pio
n contribution from 0.8 GeV 
to $\infty$. The final numbers for the two choices of $\tin$ are compiled in Table \ref{table:amudavier}.

\begin{table}
\begin{center}
\caption{ $\pi^{+}\pi^{-}$ contribution  to the muon anomaly for energies above  
$\sqrt{\tin}$.}\vspace{0.1cm}
\label{table:amudavier}
\renewcommand{\tabcolsep}{1.5pc} 
\renewcommand{\arraystretch}{1.1} 
\begin{tabular}{cc}\hline
$\sqrt{\tin}$  & $\hat a^{\pi\pi}_\mu$  \\\hline 	
$0.800\,\gev$ &  $95.23 \times 10^{-10}$  \\
$0.917\, \gev$ &   $22.17 \times 10^{-10}$ \\ 
\noalign{\smallskip}\hline
\end{tabular}
\end{center}
\end{table}

 Finally,  we use  additional spacelike data coming from  \cite{Horn,Huber}, which are
collected for completeness in Table  \ref{table:Huber}, where the first error is statistical and 
the second is systematical.

\vspace{0.5cm}
\begin{table}
\begin{center}
\caption{Spacelike data from \cite{Horn,Huber}.}
\label{table:Huber}
\renewcommand{\tabcolsep}{1.5pc} 
\renewcommand{\arraystretch}{1.1} 
\begin{tabular}{c c c}
\hline\noalign{\smallskip}
 $t$&  Value[$\gev^2$] & $F(t)$  \\
\noalign{\smallskip}\hline\noalign{\smallskip}    	
$t_1$ & $-1.60$ & $0.243 \pm  0.012_{-0.008}^{+0.019}$  \\ 
$t_2$ &$ -2.45 $ & $ 0.167 \pm 0.010_{-0.007}^{+0.013}$  \\
\noalign{\smallskip}\hline
\end{tabular}
\end{center}
\end{table}
\section{Allowed domain in the $c-d$ plane}\label{sec:cd}
In this section, we present the constraints on the coefficients $c$ and $d$ entering the Taylor expansion (\ref{eq:taylor})
using the formalism developed in Sec. \ref{sec:method}. 
We list in Table \ref{table:coefficients} the various quantities required in the 
basic inequality (\ref{eq:det}), for two choices of $\tin$. We implemented the normalization $F(0)=1$, 
but kept arbitrary the charge radius  $\la r^2_\pi \ra$ and the spacelike values $F_1$ and $F_2$. Using 
the input from Tables \ref{table:amudavier} and \ref{table:coefficients}, one obtains easily from (\ref{eq:det}) 
a convex quadratic condition for the  coefficients $c$ and $d$, represented as the interior of an ellipse 
in the $c-d$ plane.

\begin{table*}
\begin{center}
\caption{Tabulation of the quantities entering as input in (\ref{eq:det}) for obtaining the constraints 
on the $c,d$ coefficients, for two choices of $\tin$. The numbers $z_n \equiv \tilde z(t_n)$ are obtained 
using (\ref{eq:ztin}) and $t_n$ given in Table \ref{table:Huber}. The numerical coefficients include the 
information on the phase below $\tin$ and the normalization $F(0)=1$, while the charge radius $\la r^2_\pi \ra$ 
(expressed in ${\rm fm}^2$) and the values $F_n\equiv F(t_n)$ are left arbitrary. }
\label{table:coefficients}
\renewcommand{\tabcolsep}{0.9pc} 
\renewcommand{\arraystretch}{1.2} 
\begin{tabular}{lll}
\noalign{\smallskip}\hline
\hspace{-0.3cm}Quantity &  $\tin = (0.8\, \gev)^2$ &$\tin = (0.917 \,\gev)^2$ \\
\noalign{\smallskip}\hline\noalign{\smallskip} 
$g_0$  & $0.2284 \times 10^{-4}$	&$0.1238\times 10^{-4}$ \\ 
$g_1$ & $(0.2503 \la r^2_\pi \ra -0.0414) \times 10^{-3}$ & $(0.1783 \la r^2_\pi\ra-0.0431)\times 10^{-3}$ \\ 
$g_2$ & $(0.1497 c-0.9547\la r^2_\pi \ra-0.1160) \times 10^{-3}$ & $(0.1401 c-0.9773 \la r^2_\pi \ra -0.0985) \times 10^{-3}$ \\
$g_3$ & $(-0.8704 c +0.3833 d +0.3879 \la r^2_\pi \ra-0.7260)\times 10^{-3}$ &$(-1.0481c+0.4712 d+0.3589\la r^2_\pi \ra-0.9154) \times 10^{-3} $ \\
$z_1$ & -0.3033 &-0.2603\\
$z_2$ & -0.3745 &-0.3285\\
$g(z_1)$ & $F_1\times 0.3051  \times 10^{-4}$ & $F_1\times 0.2066 \times 10^{-4}$  \\ 
$g(z_2)$ & $F_2 \times 0.3984\times 10^{-4}$  & $F_2 \times 0.2210 \times 10^{-4}$\\ 
\noalign{\smallskip}\hline
\end{tabular}
\end{center}
\end{table*}

We consider first the constraints obtained without any information at spacelike points, 
when the determinant (\ref{eq:det}) has only one element,  $\bar I$, and the condition (\ref{eq:det}) becomes
\beq\label{eq:L2gi}
g_0^2+g_1^2+g_2^2+g_3^2+\ldots \leq \hat a_\mu^{\pi\pi}.
\eeq
The quantities $g_i$ are calculated for $\tin = (0.8\, \gev)^2$ using the first line of Table \ref{table:amudavier} 
and the first column of  Table \ref{table:coefficients}, and for $\tin = (0.917\, \gev)^2$  using the quantities 
written in the second line of Table \ref{table:amudavier} and the second column of  Table \ref{table:coefficients}.

\begin{figure}[htb]
\vspace{0.35cm}
 \includegraphics[width = 7.cm]{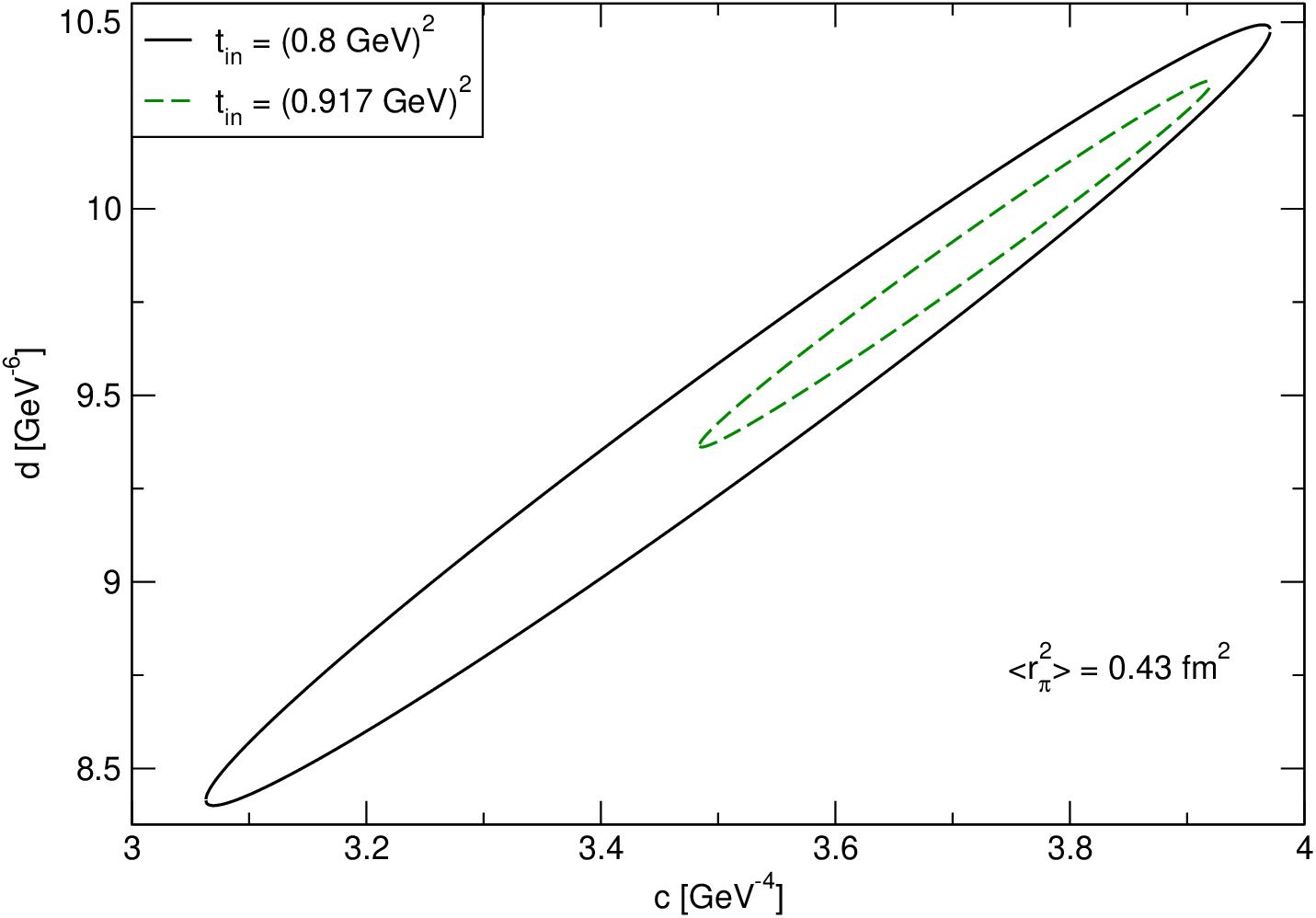}
\caption{Comparison of the $c-d$ domain obtained with  $\tin = (0.917 \, \gev)^2$ and  $\tin = (0.8 \, \gev)^2$ for  $\la r^2_\pi \ra = 0.43 \,{\rm fm}^2$.}
\label{fig:fig2}
\end{figure}

In order to investigate the influence of the choice of the threshold $\tin$, we show in Fig.\ref{fig:fig2} the 
domains obtained with the two values of $\tin$ considered in Tables \ref{table:amudavier} and \ref{table:coefficients}.  
For convenience, we take $\la r^2_\pi \ra = 0.43\, {\rm fm}^2$ \cite{CGL,TrYn1,TrYn2,PiPo}. The figure shows that the ellipse 
corresponding to $\tin = (0.917\,\gev)^2$ is smaller and lies fully inside that of the ellipse
with  $\tin = (0.8\,\gev)^2$, proving that the best results are obtained by exploiting the known phase along 
the whole elasticity region. Therefore, in what follows we shall adopt the choice $\tin = (0.917\, \gev)^2$.

A precise estimate  $\la r^2_\pi \ra = (0.435\pm 0.005)\, {\rm fm}^2$ is given in \cite{CGL}.  In Fig.\ref{fig:fig1} 
we present the domains described by (\ref{eq:det}) for $\tin = (0.917\,\gev)^2$ and two values of the charge radius
 $\la r^2_\pi \ra = 0.43\, {\rm fm}^2$ and  $\la r^2_\pi \ra = 0.44\, {\rm fm}^2$ resulting from this estimate.  
The allowed domain is quite sensitive to the variation of $\la r^2_{\pi} \ra$, being shifted towards the upper 
right end if  $\la r^2_\pi \ra$ is increased. To account for the uncertainty of the charge radius, we take as 
allowed domain the union of the two ellipses in Fig.\ref{fig:fig1}, which leads to the ranges
\bea\label{eq:cdnum1}
3.48\, \gev^{-4} \lesssim c \lesssim 3.98\, \gev^{-4}, \nonumber\\ 9.36\, \gev^{-6}\lesssim d \lesssim 10.46\, \gev^{-6}, 
\eea
with a strong correlation between the values of $c$ and $d$.

\bigskip

\begin{figure}[htb]
\vspace{0.35cm}
  \includegraphics[width = 7.cm]{Fig1.eps}
\caption{Allowed domain in the  $c-d$ plane obtained with $\tin= (0.917\,\gev)^2$,  for   $\la r^2_\pi \ra = 0.43\, {\rm fm}^2$ and   
$\la r^2_\pi \ra = 0.44\, {\rm fm}^2$.  }
\label{fig:fig1}
\end{figure}

\begin{figure}[htb]
\vspace{0.35cm}
  \includegraphics[width = 7.cm]{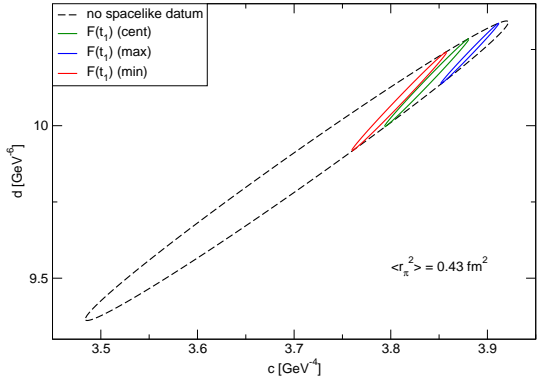}
\caption{Allowed domain in the $c-d$ plane calculated with  $\tin = (0.917 \, \gev)^2$ and $\la r^2_\pi \ra = 0.43 \,{\rm fm}^2$,  
for three values   of $F(t_1)$ at the spacelike point $t_1=-1.6\,\gev^2$ (central value in  Table \ref{table:Huber} 
and the extreme values obtained from the error intervals). Also shown is the large ellipse when no spacelike datum is included.}
\label{fig:fig3}
\end{figure}

We implement now the value at a point on the spacelike axis, using the input given in Table \ref{table:Huber}.  
In this case  the determinant in (\ref{eq:det}) has two rows and two columns. We choose the input at the 
spacelike point $t_1$ given in Table \ref{table:Huber} and account for the errors by varying  $F_1$ inside 
the error bars. In Fig. \ref{fig:fig3} we present the allowed domain in the $c-d$ plane obtained for 
$\la r^2_\pi \ra = 0.43 \,{\rm fm}^2$ and three values of  $F_1$: the central value 0.243 given in 
Table \ref{table:Huber}, and the extreme values 0.265 (0.228) obtained by adding (subtracting) the 
corresponding errors added in quadrature. The additional information on the spacelike axis improves 
in a dramatic way the constraints on the $c$ and $d$ coefficients. The small ellipses are entirely 
included in the larger ellipse obtained without information on the spacelike axis, which confirms 
the consistency of the various pieces of the input information. 
Varying  $F_1$ inside the error bars, we obtain the allowed domain of the 
$c$ and $d$ parameters  at the present level of knowledge as the union of the three  small  ellipses in Fig.  \ref{fig:fig3}.
This gives, for $\la r^2_\pi \ra = (0.435\pm 0.005) \,{\rm fm}^2$, the allowed ranges 
\bea\label{eq:cdnum2}
3.75\,\gev^{-4} \lesssim c \lesssim 3.98\,\gev^{-4},\nonumber\\  9.91\,\gev^{-6}  \lesssim d \lesssim 10.45\,\gev^{-6} , 
\eea
with a strong correlation between the two coefficients. The comparison with (\ref{eq:cdnum1}) shows that 
the information at the spacelike point improves the lower bounds on both $c$ and $d$, a feature seen actually 
from Fig.  \ref{fig:fig3}.

Similar results are obtained using as input the second Huber datum $t_2$ in Table \ref{table:Huber}. Note 
that the formalism allows the simultaneous inclusion of several spacelike points in the determinant (\ref{eq:det}). 
In practice, as discussed in \cite{AnRa1,AnRa2}, when more points are included the results are extremely 
sensitive to the values used as input, which requires adequate numerical methods for treating the problem.

To illustrate the issues that arise in this context, a small 
digression is in order: 
the formalism presented in this work can be used to obtain limits
on the value $F(t_2)$ at the second spacelike point, 
 given the value $F(t_1)$ at the first one.  This range results from the general 
inequality (\ref{eq:det}), written  as
\beq\label{eq:detz1z2}
\left|
\begin{array}{c c c }\vspace{0.2cm}
\hat{a}^{\pi\pi}_\mu-g_0^2-g_1^2  &~~~\bar\xi_1 &~~~\bar\xi_2 \\\vspace{0.2cm}	
 \bar\xi_1 &~~~ \D \frac{z^{4}_{1}}{1-z^{2}_1} &~~~\D \frac{(z_1z_2)^2}{1-z_1z_2}  \\\vspace{0.2cm}
 \bar\xi_2 &~~~\D \frac{(z_1 z_2)^{2}}{1-z_1 z_2} &~~~\D \frac{(z_2)^{4}}{1-z_2^2}  \\
\end{array}\right| \geq 0,
\eeq
where $z_i=\tilde z(t_i)$ and $\bar\xi_i=g(z_1)- g_0 -g_1 z_i$. 

Using as input the coefficients given in the second column of Table \ref{table:coefficients}, we obtain from the above inequality a strong correlation between the values $F(t_1)$ and $F(t_2)$. For instance, taking the radius to be 0.435 fm$^{2}$ and $F(t_1)$ at its central value in Table \ref{table:Huber},  (\ref{eq:detz1z2}) restricts $F(t_2)$ to  the narrow range $(0.159, 0.173)$. The central experimental value of $F(t_2)$  quoted in Table \ref{table:Huber} is contained in this range, which means that the central Huber values are consistent with each other in the analytic framework that we have adopted. On the other hand, taking $F(t_1)$  at the lower end of the experimental interval, we obtain  the allowed range of $F(t_2)$ as $(0.135, 0.153)$, below the 
experimental interval, while fixing $F(t_1)$ at the upper end yields the range $(0.199, 0.201)$, above the experimental interval. It follows that an allowed range of  $F(t_2)$ consistent with the experiment  can be obtained only by reducing  the input range of $F(t_1)$. 
By varying simultaneously the value of $F(t_1)$ and the radius, $\la r^2_\pi \ra = (0.435 \pm 0.005) \,{\rm fm}^2$, we obtain for  $F(t_2)$  the range $(0.130, 0.201)$, which may be expressed as $F(t_2)=0.166 \pm 0.036$.  The result is consistent with the numbers in Table \ref{table:Huber}, but the error is a bit larger than the actual experimental one. 

The digression above shows also that, by imposing simultaneously the experimental values at $t_1$ and $t_2$, we can only obtain a slight improvement of the allowed domain of the parameters $c$ and $d$. The reason is the fact that, as already noted above, the information on $F(t_2)$ forces $F(t_1)$ to lie within a slightly smaller  range  around the central value.  Since the gain is expected to be small, we keep for simplicity as input only one spacelike point, which is sufficient to produce the narrow ranges reported in (\ref{eq:cdnum2}). 

It is of interest to compare our predictions with previous determinations available in the literature.
First, the range of $c$ given  in (\ref{eq:cdnum2}) considerably improves the bounds obtained with  similar techniques 
in \cite{IC, AnRa1,AnRa2,Abbas:2009}. The improvement is due mainly to the very accurate information available now 
on the modulus, expressed in the values  in Table \ref{table:amudavier}.  On the theoretical side,  a fit based on ChPT to two-loop accuracy for $\tau$ decays gives  $c=(3.2 \pm 0.5_{exp} \pm 0.9_{theor})\,\gev^{-4}$  \cite{CFU}. Subsequent calculations of the electromagnetic form factors in two-loop ChPT lead to the  values $c=(3.85 \pm 0.60)\,\gev^{-4}$  \cite{BiCo}, in agreement with the range given  in (\ref{eq:cdnum2}), and $c=(4.49 \pm 0.28)\,\gev^{-4}$  \cite{BiTa}, slightly above  that range.  Finally, both the prediction $c =(4.00 \pm 0.50)\,{\rm GeV}^{-4}$, based on the quark-mass dependence of the form factor \cite{Guo}, and the  range $c=(4 \pm 2)\,\gev^{-4}$  quoted as a conservative next-to-next-to-leading ChPT result in the same reference, are consistent with  (\ref{eq:cdnum2}).  On the other hand, a recent lattice calculation with chiral extrapolation based on two-loop ChPT gives  a slightly lower value, $c = 3.22(17)(36)\, \gev^{-4}$ \cite{Aoki}.  It must be noted however, that the lattice data are generated at rather high spacelike momenta, $t\in\,(-0.3, -1.7)\,\gev^{2}$.  Therefore the extraction of the radius and the curvature can not be very precise and the corresponding uncertainties might be larger than estimated.

Other determinations of the curvature are based on fits of experimental data with specific analytic 
parametrizations of the form factor. The value $c= (3.90 \pm  0.10)\,\gev^{-4}$  was obtained in  \cite{Truo} 
by a usual dispersion relation, while a fit of the ALEPH data \cite{Aleph} on the hadronic $\tau$ decay rate 
with a Gounaris-Sakurai formula for the form factor \cite{GoSa}   gives  $c= (3.2 \pm  1.0)\,\gev^{-4}$. 
Several analyses are based on phase (Omn\`es-type) representations, with various parametrizations of the phase 
along the whole unitarity cut. Their  predictions,  like $c=(3.79 \pm 0.04)\,\gev^{-4}$ \cite{PiPo},  $c=(3.84 \pm 0.02)\,\gev^{-4}$  \cite{TrYn2} and  more recently  $c= (3.75 \pm  0.33)\,\gev^{-4}$ \cite{Guo}, are in overall agreement with  (\ref{eq:cdnum2}). We note also that the value $c = (3.30 \pm 0.03_{stat} \pm 0.33_{syst})\,\gev^{-4}$,   obtained recently  from a fit of spacelike data with Pad\'{e} approximants \cite{Masjuan}, is below our prediction (\ref{eq:cdnum2}). 
It may be worth investigating whether the fact that the unitarity cut and 
the precise data available along it are not included in this analysis  is responsible for the mismatch.

  As in the case of $c$, the range of  $d$ given  in (\ref{eq:cdnum2}) considerably improves the bounds obtained with  similar techniques in \cite{IC, AnRa1,AnRa2,Abbas:2009}. The information available in the literature on the cubic term in the Taylor expansion (\ref{eq:taylor}) is not rich. Theoretical results from ChPT and  lattice calculations are not yet available. From fits of the data, the value $d=(9.70\pm 0.40) \, {\rm GeV}^{-6}$ was obtained  by means of usual dispersion relations in \cite{Truo}, while the Taylor expansion of the  Gounaris-Sakurai parametrization \cite{Aleph}, mentioned above, leads to $d= 9.80 \, {\rm GeV}^{-6}$.  Both values are consistent with the range (\ref{eq:cdnum2}).

\section{Domain where zeros are excluded}\label{sec:zeros}
As we discussed in the Introduction (see also \cite{Abbas:2010PRD}), the formalism developed in Sec.\ref{sec:method} 
allows one to find rigorously the domain where the form factor cannot have zeros. The method amounts to testing the 
consistency of the assumption that a zero is present with the other pieces of the input.  Let us assume first that 
$F(t)$ vanishes at some real point $t_0$. From (\ref{eq:gF}) it follows that $g(z_0)=0$, where $z_0=\tilde z(t_0)$.
 We therefore include this value in the determinant (\ref{eq:det}) and test the validity of the inequality: if it 
is satisfied, a zero is possible, if it is violated, the zero is forbidden. In particular, if we use only the 
information on the normalization $F(0)=1$ and the charge radius, with no input on the spacelike axis, we obtain 
from (\ref{eq:det}) and (\ref{eq:barxi}) the following condition 
\beq\label{eq:det1}
\left|
\begin{array}{l l }\vspace{0.2cm}
 \hat{a}^{\pi\pi}_\mu \!- \!g_0^2\!-\!g_1^2 &~~~~  - \!g_0\! -\!g_1 z_0\\
	 - \!g_0\!-\!g_1 z_0 &~~~~ \D \frac{z^{4}_{0}}{1-z^{2}_0}\\
	\end{array}\right| < 0,
\eeq
for the points $z_0$ such that the form factor cannot vanish at $t_0=\tilde t(z_0)$. Here $g_0$ and $g_1$ are 
expressed  cf. Table \ref{table:coefficients} in terms of the charge radius. 

If we include in addition the value at a point $z_1=\tilde z(t_1)$, the condition reads
\beq\label{eq:det2}
\left|
\begin{array}{l l l }\vspace{0.2cm}
\hat{a}^{\pi\pi}_\mu\! -\! g_0^2\!-\!g_1^2 &~~~-\! g_0\!-\!g_1 z_0 &~~~\bar\xi_1 \\\vspace{0.2cm}	
 -\!g_0\!-\!g_1 z_0 &~~~\D \frac{z^{4}_{0}}{1-z^{2}_0} &~~~\D \frac{(z_0z_1)^2}{1-z_0z_1}  \\\vspace{0.2cm}
 \bar\xi_1 &~~~\D \frac{(z_0 z_1)^{2}}{1-z_0 z_1} &~~~\D \frac{(z_1)^{4}}{1-z_1^2}  \\
\end{array}\right| < 0,
\eeq
with $\bar\xi_1=g(z_1)- g_0 -g_1 z_1$.

\begin{figure}[htb]
\vspace{0.35cm}
\includegraphics[width = 7.1cm]{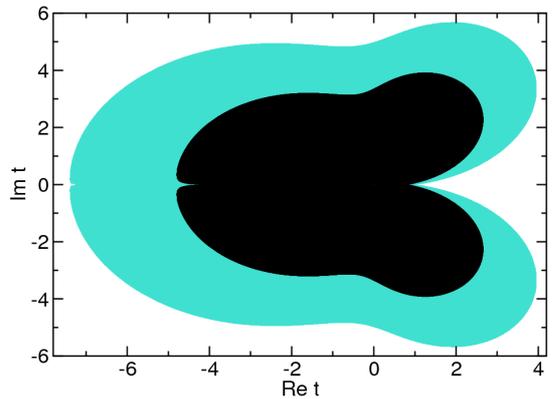}
\caption{Comparison of the domains without zeros obtained from (\ref{eq:det3}) using 
$\tin = (0.8\, \gev)^2$ (smaller domain) and  $\tin = (0.917\, \gev)^2$ (bigger domain), for  
$\la r^2_\pi \ra = 0.43\, {\rm fm}^2$.}
	\label{fig:fig6}
\end{figure}

With the values given in Tables \ref{table:amudavier} and \ref{table:coefficients} for 
$\tin = (0.917\,\gev)^2$ and $\la r^2_\pi \ra = 0.43\,{\rm fm}^2$, the inequality (\ref{eq:det1}) 
implies that simple zeros are excluded from the interval  $-1.93\,\gev^2 \leq t_0 \leq 0.83\,\gev^2$ 
of the real axis. If we  impose the additional constraint at a spacelike point $t_1=-1.6\,\gev^2$,  
the interval for the excluded zeros is much bigger. The left end of the range is actually quite 
sensitive to the input value $F_1=F(t_1)$.  Using  the central value $F_1=0.243$ given in 
Table \ref{table:Huber}, we find from (\ref{eq:det2}) that the form factor cannot  have simple 
zeros in the range $-5.56\, \gev^2 \leq t_0 \leq 0.84\, \gev^2$.  By varying $F_1$ inside the 
error interval given in Table \ref{table:Huber} (with errors added in quadrature), we find that 
zeros are excluded from the range $-12.67\, \gev^2 \leq t_0 \leq 0.84\, \gev^2$  for $F_1= 0.265 $ 
at the upper limit of the error interval, while for the lower limit $F_1=0.228$ the range 
is $-4.46\,\gev^2 \leq t_0 \leq 0.84\,\gev^2$.

\begin{figure}[htb]
\vspace{0.35cm}
 \includegraphics[width = 7.1cm]{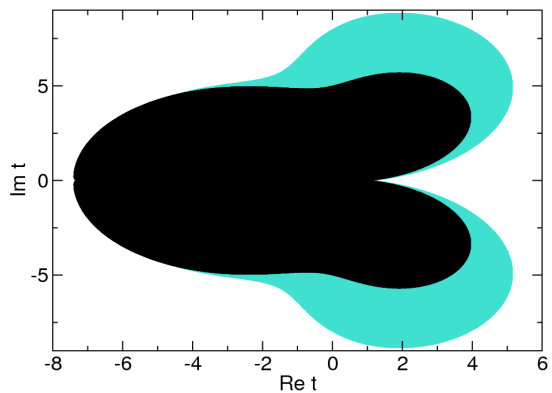}
\caption{Domain without  zeros obtained from (\ref{eq:det3}) using $\tin = (0.917\, \gev)^2$, 
for two values of the pion charge radius, $\la r^2_\pi \ra = 0.43\, {\rm fm}^2$ (smaller domain) 
and $\la r^2_\pi \ra = 0.44\, {\rm fm}^2$ (bigger domain).}
\label{fig:fig5}
\end{figure}

\begin{figure}[htb]
\vspace{0.35cm}
 \includegraphics[width = 7.1cm]{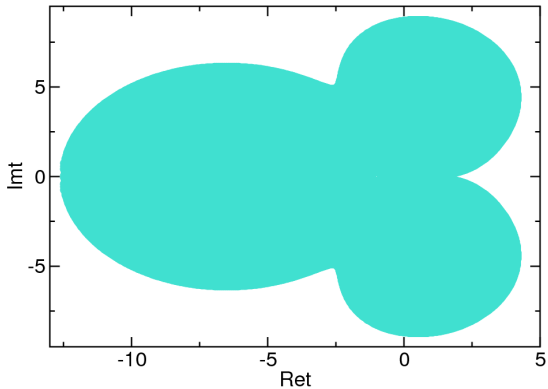}
\caption{Domain without zeros obtained  with $\tin = (0.917\, \gev)^2$ and  $\la r^2_\pi \ra = 0.43\, {\rm fm}^2$, 
using in addition   the central experimental value  $F(t_1)=0.243$ at the spacelike point $t_1=-1.6\,\gev^2$.}
	\label{fig:fig7}
\end{figure}

\begin{figure}[htb]
\vspace{0.35cm}
\includegraphics[width = 8.5cm]{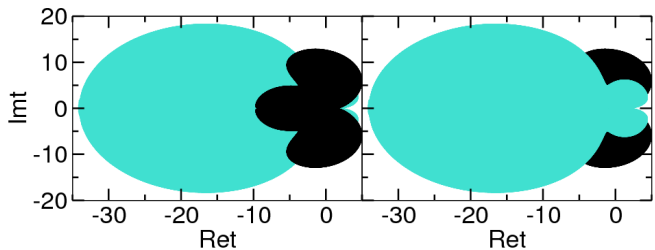}
\caption{Comparison of the domains with no zeros obtained with $\tin = (0.917\, \gev)^2$ and  
$\la r^2_\pi \ra$ = 0.43 fm$^2$, for the spacelike input  $F_1=0.265$  (bigger domain) and $F_1=0.228$  (smaller domain).}
	\label{fig:fig10}
\end{figure}

We now turn to the study of complex zeros.  The formalism presented in Sec. \ref{sec:method} can 
be easily adapted to include complex values of the form factor outside the real axis. Since the 
form factor is real analytic, its zeros occur in complex conjugate pairs, {\em i.e.} if $F(t_0)=0$, 
then also $F(t_0^*)=0$ (a double zero occurs as $t_0$ approaches the real axis). One can show that the 
determinant condition (\ref{eq:det1}) for the domain without zeros is generalized to 
\beq\label{eq:det3}
\left|
\begin{array}{l l l }\vspace{0.2cm}
\hat{a}^{\pi\pi}_\mu\! -\! g_0^2\!-\!g_1^2 &~~~-\! g_0\!-\!g_1 z_0 &~~~-\! g_0\!-\!g_1 z_0^*  \\\vspace{0.2cm}	
 -\!g_0\!-\!g_1 z_0^* &~~~\D \frac{(z_0 z_0^*)^2}{1-|z_0|^2} &~~~\D \frac{(z_0^*)^4}{1-(z_0^*)^2}  \\\vspace{0.2cm}
- \! g_0\!-\!g_1 z_0  &~~~\D \frac{z_0^4}{1-z_0^2} &~~~\D \frac{(z_0 z_0^*)^{2}}{1-|z_0|^2}  \\
\end{array}\right| < 0.
\eeq
The determinant is real since the corresponding matrix is Hermitian.
 The $4\times 4$ determinant that includes in addition a value at a spacelike point $t_1$ can be easily written down. 

 We first apply the inequality (\ref{eq:det3}) to illustrate the dependence of the domain without zeros on the 
value of $\tin$ used in the calculations.  As seen from  Fig.\ref{fig:fig6}, the larger value 
 $\tin = (0.917\, \gev)^2$ leads to a domain that extends to high values of $|t|$ in all the directions 
of the complex plane, which shows that, like in the case of the $c-d$ domain,  the best results are obtained 
if the phase condition (\ref{eq:watson}) is used along the whole range of validity. 

 The dependence of the domain on the variation of $\la r^2_\pi \ra$ is shown in Fig.\ref{fig:fig5}. As expected, 
for a larger charge radius,  $\la r^2_\pi \ra = 0.44\, {\rm fm}^2$, the zeros are excluded from  a bigger complex 
domain around the timelike axis, while the left end of the domain, around the spacelike axis, is almost insensitive 
to the slope at $t=0$.

The effect of an additional input  at a spacelike point is illustrated in Fig.\ref{fig:fig7}, where we show 
the domain in the complex plane where zeros are excluded, using  $t_1=-1.6 \gev^2$ and the value $F(t_1)=0.243$ 
(the central experimental value given in  \cite{Horn,Huber}). By comparing Fig.\ref{fig:fig7} with the large
 domain in Fig.\ref{fig:fig6}, one can see that the knowledge of the form factor at a spacelike point excludes 
zeros in a larger domain  near the spacelike axis, while it has a smaller influence on the right part of the domain. 
This feature is present also in Fig.\ref{fig:fig10}, which shows the sensitivity of the domain to the input 
value of $F(t_1)$.  As is seen in the figure, the larger value  $F(t_1)=0.265$ obtained from the upper limit 
of the error bar, excludes the zeros in a domain extending to considerably larger values along the spacelike axis.

The results on the zeros reported in the literature \cite{Cronstrom,Raszillier,RaSc,RaSc2, DuMe,Leutwyler:2002} 
are rather controversial. The best results for the regions free of zeros were obtained in \cite{Raszillier,RaSc,RaSc2}, 
by a method related to ours. However, since the  experimental information at that time was poor, the authors 
were forced to make some ad-hoc assumptions, especially on the modulus on the timelike axis.  At present the 
precise measurement of the modulus gives a solid basis to our results.

The issue of zeros is of relevance for the analytic representation of the
 form factor using  phase (Omn\`es)-  or modulus-type  representations, which require the knowledge of the zeros. 
Such representations were extensively studied in the past \cite{LaSt, GuPi, PiPo, Leutwyler:2002, Guo, Geshkenbein}, 
and  often are based on the assumption that zeros are  absent. Our results, which show that the zeros are excluded 
from a rather large region at low energies, give support to such representations, and confirm also theoretical 
expectations based on ChPT or more general physical arguments \cite{Leutwyler:2002}.

\section{Discussions and Conclusion}\label{sec:conclusion}
The experimental information available at present on the pion electromagnetic form factor is very rich. 
The recent high statistics measurements of the modulus by BABAR and KLOE collaborations \cite{BABAR,KLOE1,KLOE2}, 
supplemented by the phase  in the elastic region known with accuracy from the $P$-wave of $\pi\pi$ scattering 
\cite{ACGL,CGL,KPY}, are expected to considerably constrain the behavior on the timelike axis. The values 
of the form factor on the spacelike axis are also measured  with increasing precison. Theoretically, predictions 
on the pion form factor at low energies are available from ChPT and lattice QCD, while perturbative QCD predicts 
the behavior at high energies along the spacelike axis. The  transition to the perturbative regime is known to 
be an open problem that deserves further study in the case of the pion form factor.

Analyticity is the ideal tool for connecting the low- and high-energy regimes for physical quantities like 
the pion form factor.  The full treatment of the present rich experimental and theoretical input, which might 
overconstrain the system, is a challenge for the future investigations based on analyticity. In the present 
work we do not perform such a complete analysis, but exploit only in part the present information on the 
modulus on the unitarity cut. However, even in this limited frame we obtain quite stringent conclusions on 
the low energy properties of the form factor.  

 The conditions used as input in our approach are expressed by the phase condition (\ref{eq:watson}) and 
the integral of the modulus squared (\ref{eq:L2}), which we  further restricted by choosing the weight  
$\rho(t)$ as the kernel relevant for the two-pion contribution to the muon anomaly, cf. (\ref{eq:amu}) and 
(\ref{eq:amuI}). A more general class of suitable weights will be investigated in a future work.   Once the input is chosen, it is exploited in an optimal way by a mathematical formalism presented in Sec. \ref{sec:method},
 leading to strong correlations between the coefficients of the Taylor expansion (\ref{eq:taylor}) at $t=0$ and 
the  values of the form factor on the spacelike axis.

 Our basic results are contained in Eqs. (\ref{eq:det}) and (\ref{eq:barxi}), where the input quantities are 
defined in Tables  \ref{table:amudavier}-\ref{table:coefficients}. The numerical coefficients in Table 
\ref{table:coefficients} depend on the  normalization $F(0)=1$ and the phase of the form factor below the 
inelastic threshold $\tin$, being vary stable with respect to small variations of the phase. Moreover, 
as emphasized in Sec. \ref{sec:method}, the results are independent of the unknown phase of the form factor 
above the inelastic threshold $\tin$.  In Table \ref{table:coefficients}, the charge radius $\la r^2_\pi \ra$, 
the higher-order Taylor coefficients $c$ and $d$, and the values of the form factor at several spacelike points 
are kept free, so the formalism can be easily applied for finding model independent correlations between the  
values of the form factor at different points  and for testing the consistency of input values known from 
different sources.  

In Sec. \ref{sec:cd} we derived stringent constraints on the allowed values of the higher-order coefficients 
$c$ and $d$ of the Taylor expansion (\ref{eq:taylor}). The best results are obtained with $\tin=(0.917\,\gev)^2$, 
which corresponds to the physical inelastic threshold produced by the $\omega\pi$ channel. The charge radius  and 
an additional information at a spacelike point were used as input. In (\ref{eq:cdnum1}) and (\ref{eq:cdnum2}) 
and in Figs.\ref{fig:fig2} - Fig. \ref{fig:fig3} we illustrated the results for $\la r^2_\pi \ra$ in the 
range $(0.43-0.44)\, {\rm fm}^2$ and $F(-1.6\,\gev^2)= 0.243 \pm  0.012_{-0.008}^{+0.019}$ \cite{Horn,Huber}. 
 It is remarkable that the allowed ranges are already comparable in precision with other determinations in the 
literature  based on specific parametrizations. 

The present method can be used also to obtain bounds on the values of the form factor along the spacelike axis, using as input the information on the timelike axis, together with some values inside the analyticity domain. As discussed in Sec. \ref{sec:cd}, using as input the value $F(t_1)$ at the first Huber point, we obtain stringent limits on the value $F(t_2)$ at the second point, with a strong correlation between the two. 
Of course, the  method can be applied in principle also to  higher spacelike energies. However,
with our choice of the  weight $\rho$, we expect that the predictions will become gradually 
weaker when the energy is increased.
Indeed, since $\rho$ decreases rapidly at large momenta,  the condition (\ref{eq:L2})  provides stringent
constraints on the low energy parameters like $c$ and $d$,  but in the same time it imposes weak constraints on the behavior of the form factor at large energies. A different choice of  $\rho$ could lead to interesting
results also for the behavior at higher energies,  but this is beyond the scope
of the present analysis and will be investigated in a future work.

In   Sec. \ref{sec:zeros} we showed that the same formalism leads to an analytic description for the 
regions of the complex plane where the zeros of the form factor are forbidden.  Our results are contained 
in Eqs. (\ref{eq:det1})-(\ref{eq:det3}) and are illustrated in Figs. (\ref{fig:fig6}) -  (\ref{fig:fig10}), 
for the same input   $\la r^2_\pi \ra$ and $F(t_1)$. We obtain a rather large domain where zeros are excluded, 
which gives support to  Omn\`es-type  representations, which often assume the absence of the zeros. Our results 
also confirm  theoretical expectations on the absence of zeros at low energy, based on ChPT or general 
physical arguments \cite{Leutwyler:2002}. We note that by our method we can find rigorously the domains 
free of zeros, but we can say nothing about the remaining domains, where zeros may be present or absent. 
Alternative methods, based on modulus representations \cite{Cronstrom, Geshkenbein,Leutwyler:2002}, can rule 
out in principle the zeros from the whole complex plane provided they are absent. However,
these methods are very sensitive to the input and led to 
controversial results in the past. An update of such analyses using the recent precise determination 
of the modulus would be of much interest.

We finally note that the mathematical formalism applied in this paper may be useful also for finding 
an analytic parametrization of the form factor suitable for fitting the rich amount of experimental data. 
Namely, the representation of $F(t)$ that results from (\ref{eq:gF}) involves the known functions 
$w(z)$, which accounts for the weight $\rho(t)$, $\omega(z)$ and ${\cal O}(t)$, which implement the phase below $\tin$, and the arbitrary function 
$g(z)$, analytic in the $t$-plane cut for $t>\tin$, or equivalently in the unit disc $|z|<1$ of the $z$-plane 
defined by  the conformal mapping (\ref{eq:ztin}). The expansion (\ref{eq:gz}) is convergent in $|z|<1$, and 
moreover the coefficients satisfy the inequality (\ref{eq:L2gi}), which is very useful in order to control the 
higher orders of the expansion and the truncation error.

\vskip0.2cm
\noindent{\bf Acknowledgement:} 
BA thanks the Department of Science and Technology, the Government of India, and the Homi Bhabha Fellowships Council for 
support. IC acknowledges support from  CNCSIS in the Program Idei, Contract No.
464/2009.  We thank B. Malaescu, G. Colangelo, M. Passera and S. Ramanan for useful correspondence.

\end{document}